\documentclass{webofc}
\usepackage[varg]{txfonts}   
\begin{document}
\title{Role of repulsive interactions in the interplay with missing strange resonances}
%
%

\author{\firstname{Paolo} \lastname{Alba}\inst{1}\fnsep\thanks{\email{alba@fias.uni-frankfurt.de}}
}

\institute{Frankfurt Institute for Advanced Studies, Goethe Universit\"at Frankfurt, D-60438 Frankfurt am Main, Germany 
          }

\abstract{
The standard implementation of the HRG model has been shown to be unable to describe all the available data on QCD matter. Here we show the balance of repulsive and attractive hadronic interactions on QCD thermodynamics through observables both calculated by lattice simulations and measured in experiment. Attractive interactions are mediated by resonance formation, which are here implemented through extra states predicted by the Quark Model, while repulsive interactions are modelled by means of Excluded Volume (EV) effects. Informations on flavour dependent effective sizes are extracted. It is found that EV effects are present in lattice QCD thermodynamics, and are essential for a comprehensive description of higher order fluctuations of conserved charges.
}
\maketitle
\section{Introduction}
\label{intro}
The current abundance of results from lattice simulations \cite{Bellwied:2015lba,Bazavov:2017dus} allows for a detailed study of QCD thermodynamics. It has been shown how the standard implementation of the Hadron-Resonance Gas (HRG) model is unable to describe all the available data, with quite strong hints for the presence of extra higher-mass resonances \cite{Bazavov:2014xya,Alba:2017mqu}, being them compatible with a prediction from the Quark-Model (QM) \cite{Ebert:2009ub,Capstick:1986bm}.
On the other hand the relevance of repulsive interactions has been pointed out through the S-matrix approach \cite{Broniowski:2015oha}, and a clear connection between the experimental phase shift and fluctuations of conserved charges has been established \cite{Huovinen:2017ogf}. These interactions can be modelled in the HRG framework through the assumption of hadrons as hard spheres, which then acquire an effective radius $r$ \cite{Rischke:1991ke}.
In the present paper we study the combination of QM states with Excluded Volume (EV) effects, extracting from lattice QCD thermodynamics informations about the effective sizes of hadrons as it was already done for the pure gauge sector \cite{Alba:2016fku}. Interesting implications have been found at imaginary chemical potential \cite{Vovchenko:2017xad}.

\section{The HRG model}
\label{sec-1}

The HRG model describes a system of interacting hadrons as a system of non interacting hadrons and resonances, where resonances formation mediates attractive hadronic interactions. Thus the partition function can be written as:
\begin{equation}
\ln\mathcal{Z}(T,\{{\mu_B,\mu_Q,\mu_S}\})=\sum_{i \in Particles}(-1)^{{B_i}+1}\frac{{d_i}}{(2\pi^3)}\int d^3\vec{p}\,\ln\left[1+(-1)^{{B_i}+1}e^{-(\sqrt{\vec{p}^2+{m_i}^2}-{\mu_i})/T}\right]\,,\,\,\,
\end{equation}
where the sum runs over all the particles listed by Particle Data Group (PDG) \cite{Patrignani:2016xqp} with their properties. From this, fluctuations of conserved charges can be calculated as:
\begin{equation}
\chi^{BQS}_{lmn}=\frac{\partial^{l+m+n}(\ln\mathcal Z/T^3)}{\partial(\mu_B/T)^l\partial(\mu_Q/T)^m\partial(\mu_S/T)^n}.
\end{equation}
For details please see \cite{Alba:2014eba,Alba:2015iva}.

\subsection{Excluded Volume}
The effective radius for hadrons results into a shifted single-particle chemical potential:
\begin{equation}
\mu_i^*=\mu_i-v_i\,p\,\,,
\end{equation}
which implies a transcendental equation for pressure $p$, where $v_i=\frac{16}{3}\pi\,r_i^3$ is the single particle \emph{eigenvolume}.
In order to be consistent with the virial expansion to second order one must use, instead of eigenvolumes, the coefficients $b_{ij}=\frac{2}{3}\pi\left(r_i+r_j\right)^3$ which properly account for two particle interactions (\emph{crossterms}). This leads to a further modification of the single particle chemical potential, and to a system of coupled transcendental equations. For details see \cite{Vovchenko:2016ebv,Satarov:2016peb}.

\begin{figure*}[b]
\centering
\includegraphics[width=0.4\textwidth]{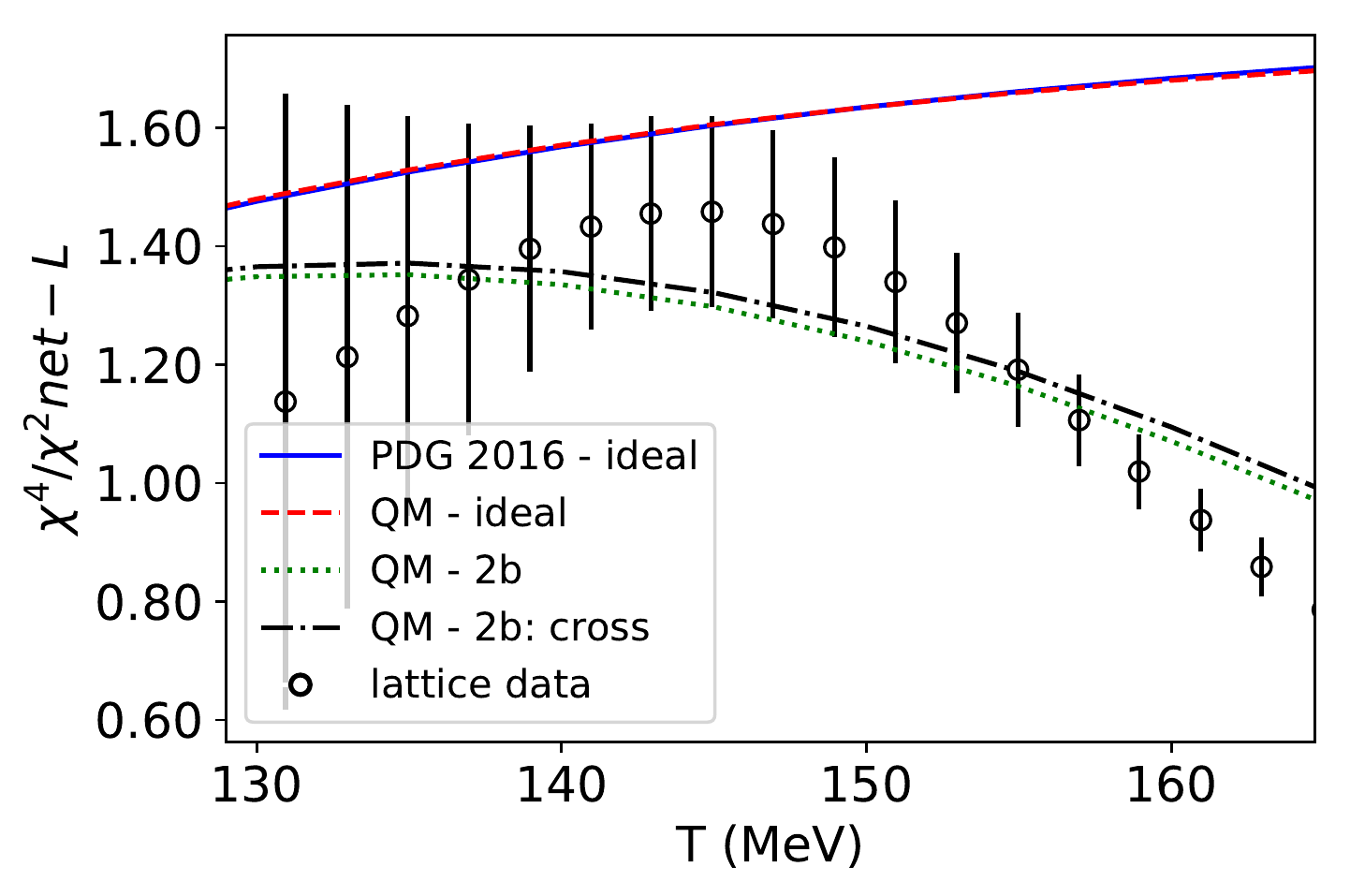}~~\includegraphics[width=0.4\textwidth]{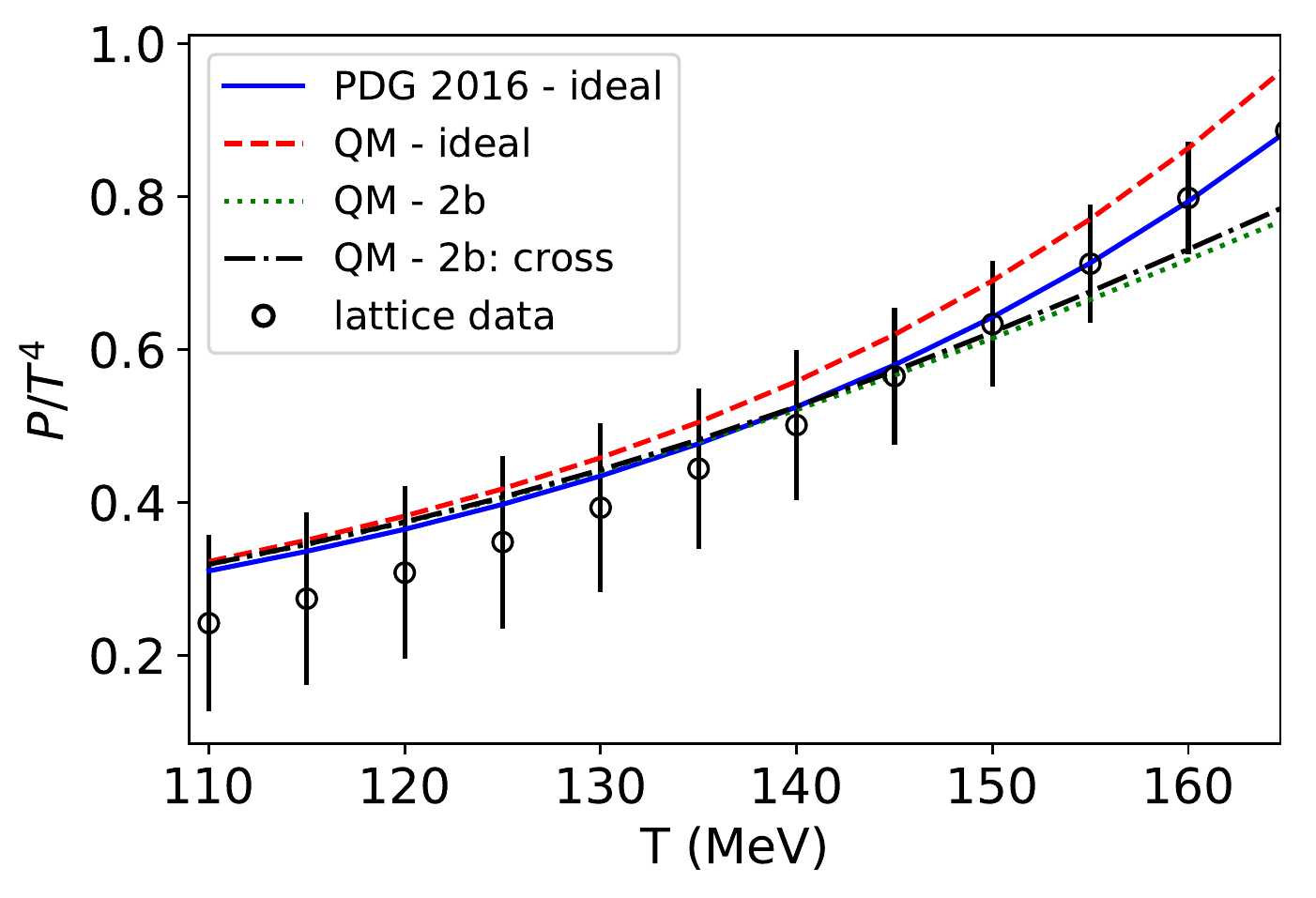}
\caption{Lattice data for $\chi^4/\chi^2$ for net-light number and pressure \cite{Bellwied:2013cta,Borsanyi:2013bia} compared to HRG calculations with different lists and EV schemes.}
\label{fig-1}
\end{figure*}

\section{Results}
We performed fits to continuum extrapolated lattice QCD data at physical values of quark masses, employing different hadronic spectra and different EV schemes. We find that the best description is given by a list inspired by QM predictions where particle eigenvolume increases with the mass, but with smaller values for strange particles with respect to light ones with equal masses (\emph{2b} scheme).\\
In figure~\ref{fig-1} it is shown how EV effects are able to reproduce the characteristic suppression of 4th to 2nd ratios of fluctuations of conserved charges, preserving the agreement with lower order observables like the pressure. The very same happens when crossterms are employed, showing that the standard EV implementation is already sufficient to catch all the relevant features of lattice QCD thermodynamics.\\
For more details see \cite{Alba:2017bbr}.

\subsection{Comparison with the experimental results}
\begin{figure}[b]
\centering
\includegraphics[width=0.5\textwidth]{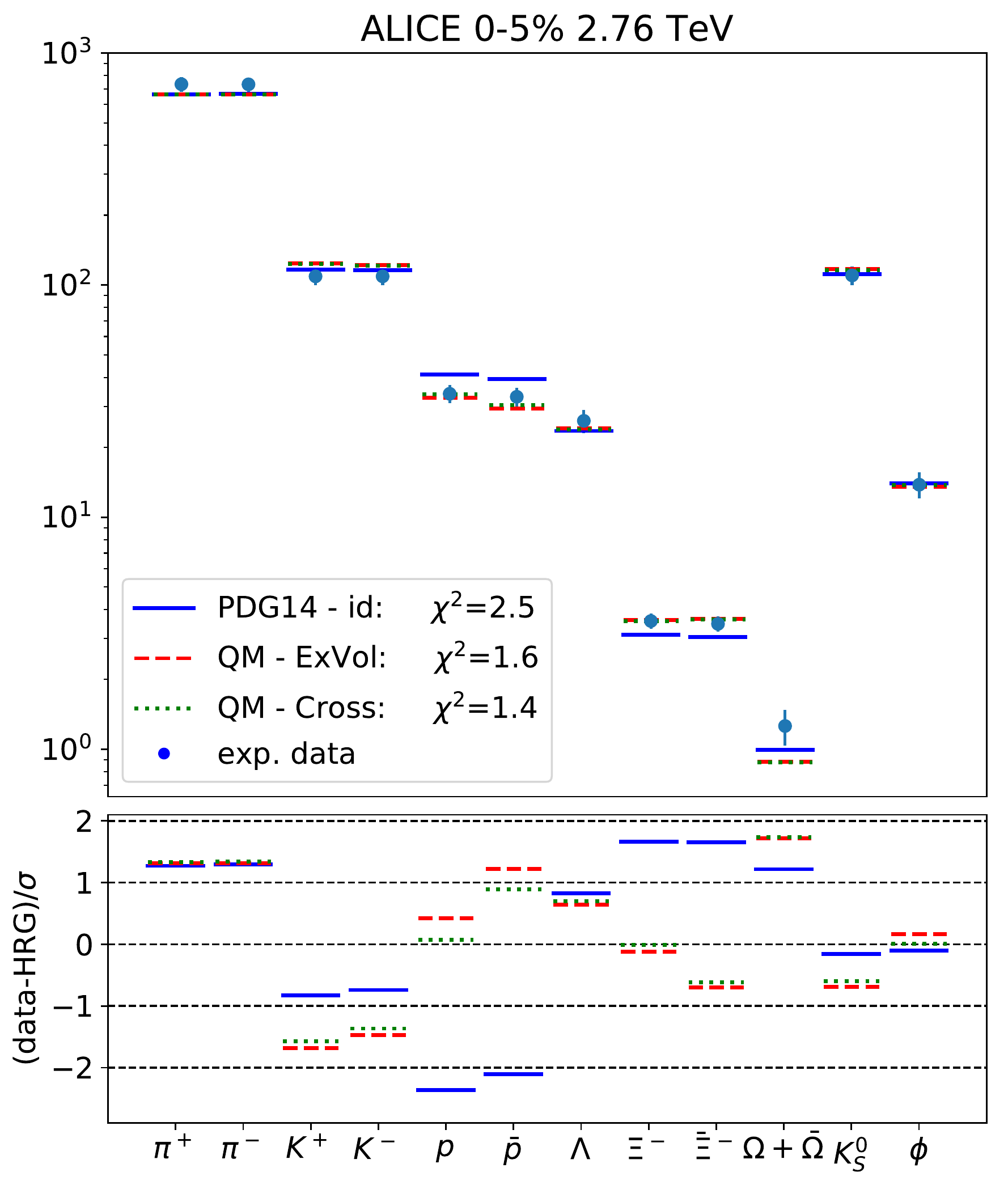}
\caption{Particle yields measured by the ALICE Collaboration for PbPb collisions at 2.76 TeV in the 0-5\% centrality class, \cite{Abelev:2013vea,Abelev:2013xaa,ABELEV:2013zaa,Becattini:2014hla} in comparison to HRG fits with different lists and EV schemes.}
\label{fig-2}
\end{figure}

In figure~\ref{fig-2} are shown fits to experimental particle yields measured by the ALICE Collaboration at 2.76 TeV $\sqrt{s_{NN}}$ in 0-5\% centrality class \cite{Abelev:2013vea,Abelev:2013xaa,ABELEV:2013zaa,Becattini:2014hla}; fit results are summarised in table~\ref{tab-1}. It is shown how the parameterisation extracted by lattice QCD essentially improves the fit mainly due to EV effects acting on (anti-)protons and (anti-)$\Xi$s, solving then the so-called proton anomaly; in particular with it there is no presence for the second minimum structure which affects some EV schemes \cite{Vovchenko:2015cbk}. This effect has already been observed in \cite{Alba:2016hwx}, where however a different scheme was employed. It should be noted that the effectiveness of the 2b scheme found here, is also supported by experimental measurement on the electric radii of a few ground-state hadrons \cite{Patrignani:2016xqp}.\\
It is worth to note that accounting for crossterms, without changing EV parameterisation, pushes the agreement with the experimental results. For more details see \cite{Alba:2017bbr}.\\

\begin{table}[b]
\centering
\caption{Parameters resulting from the fit to particle yields measured by the ALICE Collaboration, using different lists and EV schemes.}
\label{tab-1} 
\begin{tabular}{|c|c|c|c|c|c|}
\hline
 & $\chi^2$& T (MeV)& $\mu_B$ (MeV)& V (fm$^3$)\\\hline
PDG14 - id &22.24/9& 154.1 $\pm$ 2.2 & 3.5 $\pm$ 6.9 & 4992.9 $\pm$ 643.7 \\
QM - ExVol &14.49/9& 149.1 $\pm$ 1.7 & 7.7 $\pm$ 7.9 & 7560.9 $\pm$ 697.2 \\
QM - Cross &12.95/9& 149.4 $\pm$ 1.7 & 7.5 $\pm$ 7.5 & 7274.8 $\pm$ 698.6 \\\hline
\end{tabular}
\end{table}

\section{Conclusions}
In this proceeding we show how EV effects are manifest both in lattice QCD thermodynamics and in experimentally measured particle yields. This was achieved without the inclusion of additional van-der-Waals attractive terms \cite{Vovchenko:2016rkn}, which may attain rather large values if observables on the lattice are fitted \cite{Samanta:2017yhh}. Furthermore in our approach we can consistently assign a finite radius to mesons too, fact that could have interesting implications on key observables. See \cite{Alba:2017bbr} for more details.\\

\vspace{-0.19cm}

\bibliography{cites}{}

\end{document}